\documentclass[twocolumn,showpacs,pra,aps,amssymb,superscriptaddress]{revtex4}
\usepackage{bm}
\usepackage{color}
\usepackage{amsmath}
\usepackage{amsthm}
\usepackage{amsfonts}
\usepackage{bm}
\usepackage[dvips]{graphicx}
\usepackage{psfrag}
\newcommand{\La}{\Lambda}
\newcommand{\bs}{\backslash}
\newcommand{\vac}{\Phi_{\rm vac}}
\newcommand{\lrarrow}{\leftrightarrow}
\newcommand{\aki}[1]{{\bf #1}}
\newcommand{\inprod}[3]{\langle #1,#2 #3 \rangle}
\newcommand{\PhiG}{\Phi_\mathrm{G}}
\newtheorem{thm}{Theorem}

\begin{document}

\title{
Nagaoka states in the SU($n$) Hubbard model
}
\author{Hosho Katsura}
\affiliation{Department of Physics, Gakushuin University, Toshima-ku, Tokyo 171-8588, Japan}
\author{Akinori Tanaka}
\affiliation{Department of General Education, Ariake National College of Technology, Omuta
836-8585, Japan}

\date{\today}

\begin{abstract}
We present an extension of Nagaoka's theorem in the SU($n$) generalization of the infinite-$U$ Hubbard model. 
It is shown that, when there is exactly one hole, the fully polarized states analogous to the ferromagnetic states in the SU($2$) Hubbard model are ground states. 
For a restricted class of models satisfying the connectivity condition, these fully polarized states are the unique ground states up to the trivial degeneracy due to the SU($n$) symmetry. 
We also give examples of lattices in which the connectivity condition can be verified explicitly. The examples include the triangular, kagome, and 
hypercubic lattices in $d$ $(\ge 2)$ dimensions, among which the cases of $d=2$ and $3$ are experimentally realizable 
in ultracold atomic gases loaded into optical lattices. 
\end{abstract}


\maketitle

\section{Introduction}
\label{sec:intro}

The (fermionic) Hubbard model has attracted considerable attention 
as a model for describing correlated electrons in solids. 
Despite its apparent simplicity, it is difficult to analyze the model in a consistent way, and exact and/or rigorous results are very limited~\cite{Lieb_review_1995, Tasaki_review_1998, Essler_textbook}. 
A number of approximate treatments have been developed to understand the model. 
A systematic approach from the large-$n$ limit of the SU($n$) Hubbard model was initiated by Affleck and Marston 
in the context of high-temperature superconductivity~\cite{Affleck_Marston_1988}. 
In this approach, the SU($n$) symmetry used is just a theoretical tool rather than a physical reality. 
Therefore, the model at finite $n$ did not attract much attention and was much less studied.  
However, a realization of the SU($n$) Hubbard model~\cite{Honerkamp_2004} and its two-orbital generalization~\cite{Gorshkov_2009} have recently been proposed theoretically in the context of ultracold fermionic atoms.  
These systems were later realized experimentally using ultracold Yb atoms~\cite{Takahashi_group_2012, Takahashi_group_2010}. 
The enlarged symmetry of the system results from the hyperfine spin degrees of freedom. 
In the large-$U$ limit, these models with $m$ ($< n$) atoms per site reduce to SU($n$) spin Hamiltonians in which a variety of exotic states including chiral spin liquids are found to be ground states~\cite{Hermele_2009, Hermele_2011}.

In this paper, we present an extension of Nagaoka's theorem~\cite{Nagaoka_1966}. 
The original theorem is the first rigorous result about the ferromagnetism in the Hubbard model. 
Nagaoka proved that when there is exactly one hole, 
the ferromagnetic state is the ground state of the infinite-$U$ Hubbard model 
if the lattice satisfies certain connectivity condition. 
A generalized version of this theorem with a simplified proof was given by Tasaki~\cite{Tasaki_1989}.
A natural analogue of the Nagaoka ferromagnetic state in a system with SU($n$) symmetry is 
a fully polarized state characterized by the Young tableau having one row with $N_{\rm f}$ boxes, 
where $N_{\rm f}$ is the total number of fermions. 
In this paper, we show that an analogue of Nagaoka's theorem holds in the SU($n$) Hubbard model 
and those fully polarized states are indeed the ground states 
if certain conditions are satisfied. 
Note that the analogue of the Nagaoka state in the SU(4) Hubbard model on small clusters 
was studied in Ref.~\cite{Miyashita_2009} and, very recently, 
the existence of fully polarized phase in the SU(3) case at certain fillings 
was indicated in Ref.~\cite{Rapp_2011}.


The rest of the paper is organized as follows. 
In Sec. \ref{sec:main}, we give a precise definition of the model and 
describe the symmetries of the Hamiltonian. 
Our main results are then presented as theorems. 
In Sec. \ref{sec:proof}, we first construct a basis in which all of the off-diagonal elements of 
the Hamiltonian are nonpositive. Then we give the definition of the connectivity condition 
and give proofs of the theorems. 
In Sec. \ref{sec:examples}, we give several examples of lattices 
that satisfy the connectivity condition. 
We conclude with a summary and outlook in Sec. \ref{sec:summary}.

\section{The model and the results}
\label{sec:main}
\subsection{The Hamiltonian}
We begin with the definition of the SU($n$) Hubbard model. 
Let $\Lambda$ be a finite lattice of $N_{\rm s}$ sites. 
The model is described by the following Hamiltonian:
\begin{eqnarray}
\label{eq:Ham}
H &=& \sum_{x,y\in \Lambda} \sum^n_{\alpha=1} t_{x,y} c^\dagger_{x,\alpha} c_{y,\alpha}
+ V( \{ n_x \})         \nonumber \\
&& 	   +\frac{U}{2} \sum_{x \in \Lambda} n_x (n_x-1),  
\label{eq:ham}
\end{eqnarray}
where $c^\dagger_{x, \alpha}$ ($c_{x,\alpha}$) creates (annihilates) 
a fermion with flavor $\alpha$ at site $x$, and the number of fermions at site $x$ 
is defined by $n_x = \sum^n_{\alpha=1} n_{x, \alpha}$ with $n_{x,\alpha}=c^\dagger_{x,\alpha} c_{x,\alpha}$. 
The hopping matrix elements are arbitrary as long as they are real,
$t_{x,x}=0$, and $t_{x,y}=t_{y,x} \ge 0$. 
Although this requirement might look odd, $t_{x,y} \le 0$ can be achieved, 
in the case of bipartite lattices,
by local gauge transformations 
($c^{(\dagger)}_{x,\alpha} \to -c^{(\dagger)}_{x,\alpha}$) 
for one sublattice. 
Furthermore, it has recently been proposed that the sign change of the hopping matrix elements 
can be achieved by shaking optical lattices~\cite{Eckardt_2010, Koghee_2011}. 
The second term $V(\{ n_x \})$ is an arbitrary real valued function of the number operators $n_{x}$. 
Typical examples are on-site potentials and charge-charge interactions between any pair of sites. 
In cold atom systems, an external harmonic trap is usually unavoidable. 
Such a confinement term can be taken into account by setting
\begin{equation}
V(\{ n_x \}) = \sum_{x\in V} \frac{1}{2} m \omega^2 |{\bm R}_x|^2 n_{x}, 
\label{eq:trap}
\end{equation}
where $m$ is the mass of atoms, $\omega$ is the trapping frequency, and 
${\bm R}_x$ is the position vector for the site $x$. 
Since we are concerned with the model on finite lattices, 
the expectation value of $V(\{ n_x \})$ in Eq. (\ref{eq:trap}) in any state 
is finite provided that $m \omega^2 < \infty$. 
The third term in Eq. (\ref{eq:ham}) represents the on-site Coulomb repulsion ($U>0$).

\subsection{Symmetries of the Hamiltonian}
Let us first consider symmetries of the Hamiltonian $H$. 
In addition to the trivial conservation of the total number of fermions, 
the Hamiltonian exhibits U($n$) $=$ U(1)$\times$SU($n$) symmetry. 
To see this, we define a set of number operators and flavor-raising, 
lowering operators as 
\begin{equation}
F^{\alpha, \alpha} = \sum_{x \in \Lambda} n_{x,\alpha},~~~
F^{\alpha, \beta} = \sum_{x \in \Lambda} c^\dagger_{x,\alpha} c_{x, \beta}.
\label{eq:gene}
\end{equation}
They satisfy the commutation relations 
$[F^{\alpha, \beta}, F^{\gamma, \delta}] 
= \delta_{\beta,\gamma} F^{\alpha,\delta} - \delta_{\delta,\alpha} F^{\gamma,\beta}$, 
which follows from the relation
\begin{equation}
[ c^\dagger_{x, \alpha} c_{x,\beta}, c^\dagger_{y,\gamma} c_{y, \delta} ] 
=  \delta_{x,y} (\delta_{\beta,\gamma} c^\dagger_{x, \alpha} c_{x, \delta} 
-  \delta_{\alpha, \delta} c^\dagger_{x, \gamma} c_{x, \beta}).
\end{equation} 
Using a similar relation, one can confirm that 
$[ F^{\alpha,\beta},n_x]=0$ and $[ F^{\alpha,\beta},
\sum^n_{\gamma=1} c_{x,\gamma}^\dagger c_{y,\gamma}]=0$, 
and thus the Hamiltonian $H$ commutes with $F^{\alpha,\beta}$, i.e., $[H, F^{\alpha,\beta}]=0$.  
From the operators $F^{\alpha,\beta}$, one can construct new operators 
that also commute with the Hamiltonian: 
\begin{eqnarray}
N &=& \sum_{x \in \Lambda} n_{x}, 
\\
T^a &=& \sum_{x \in \Lambda} \sum_{\alpha,\beta} c^\dagger_{x,\alpha} {\cal T}^a_{\alpha,\beta} c_{x, \beta},~~
(a=1,...,n^2-1),
\end{eqnarray}
where ${\cal T}^a_{\alpha,\beta}$ are the generators of SU($n$) Lie algebra. 
Therefore it is concluded that the Hamiltonian has a global U($n$) $=$ U(1)$\times$SU($n$) symmetry. 
As a side remark, we note that the restriction of the Hamiltonian to the subspace 
where each site is occupied by at most two fermions 
has an enhanced symmetry~\cite{Ying_2001}. 
Moreover, this projected model in one dimension is exactly solvable 
by means of the Bethe ansatz~\cite{Choy_Haldane_1982, Lee_Schlottmann_1989, Frahm_1993}.

\subsection{Fully polarized states}
\label{sec: full pol}
In this subsection, we introduce the notion of fully polarized states in the SU($n$) Hubbard model. 
Since the numbers of $\alpha$ fermions, $F^{\alpha,\alpha}$, are conserved, 
the eigenstates of $H$ are separated into disconnected sectors labeled by the eigenvalues of $F^{\alpha,\alpha}$ ($\alpha=1, ..., n$). 
In what follows, we shall denote eigenvalues of $F^{\alpha,\alpha}$ by $N_{\alpha}$.  
The off-diagonal operators $F^{\alpha,\beta}$ ($\alpha \ne \beta$) play a role 
in connecting degenerate eigenstates in different subspaces. 
For instance, starting from a fully polarized state $\Phi_1$ that is an eigenstate of $H$ 
in the subspace $N_1=N_{\rm f}, N_2 = \cdots N_n=0$, one can obtain degenerate states with the same energy 
by applying $F^{\alpha,\beta}$ repeatedly: 
\begin{equation}
\Phi^{(N_1,...,N_n)}_1 = (F^{n,1})^{N_n}\cdots (F^{3,1})^{N_3} (F^{2,1})^{N_2}\Phi_{1},
\label{eq:pol states}
\end{equation}
where $N_1 = N_{\rm f}-\sum^{n}_{\alpha=2} N_\alpha$. 
Note that we have assumed that $\sum^{n}_{\alpha=2} N_\alpha \le N_{\rm f}$, 
where $N_{\rm f}$ is the total number of fermions. 

To show that the above state is indeed the eigenstate, 
we shall prove by induction that the squared norm of $\Phi^{(N_1,...,N_n)}_1$ is nonvanishing. 
We first set $N_2=\cdots=N_n=0$. Then the statement is trivial because $|\Phi_1|^2 > 0$. 
Next, we suppose that the squared norm of the state
\begin{equation}
\Phi^{(N_1, ..., N_k)}_1 = (F^{k,1})^{N_k}\cdots (F^{3,1})^{N_3} (F^{2,1})^{N_2}\Phi_{1},
\end{equation}
is nonvanishing. Then from the commutation relation $[F^{1,k+1}, F^{k+1,1}] = F^{1,1} - F^{k+1,k+1}$ 
and the fact that $F^{1,k+1} \Phi^{(N_1, ..., N_k)}_1= 0$, we have 
\begin{eqnarray}
&& F^{1,k+1} (F^{k+1, 1})^{m} \Phi^{(N_1, ..., N_k)}_1 \nonumber \\
&=& m (N_1 -m+1)  (F^{k+1, 1})^{m-1} \Phi^{(N_1, ..., N_k)}_1 
\end{eqnarray}
for $m=1,2,...,N_{k+1}$. Using the above chain of relations, we obtain 
\begin{eqnarray}
&& |\Phi^{(N_1, ..., N_k, N_{k+1})}_1|^2 
=  | (F^{k+1,1})^{N_{k+1}} \Phi^{(N_1, ..., N_k)}_1 |^2 \nonumber \\
&=& \frac{N_1 ! N_{k+1}!}{(N_1-N_{k+1})!} |\Phi^{(N_1, ..., N_k)}_1|^2,
\end{eqnarray}
and find that the squared norm of $\Phi^{(N_1, ..., N_{k+1})}_1$ is nonvanishing when $N_{k+1} \le N_1$. 
The desired result $|\Phi^{(N_1, ..., N_{n})}_1|^2 > 0$ then follows by induction.

The above argument ensures that the number of the states with the same energy as $\Phi_1$ is at least
\begin{equation}
d_{\rm deg} = \binom{N_{\rm f}+n-1}{N_{\rm f}} = \frac{(N_{\rm f}+n-1)!}{N_{\rm f}!(n-1)!}. 
\label{eq:deg}
\end{equation}
This number coincides with the number of standard Young tableaux 
having one row with $N_{\rm f}$ boxes. When $n=2$, $d_{\rm deg}=N_{\rm f}+1$, which is the number of ferromagnetic states in the SU(2) Hubbard model with $N_{\rm f}$ fermions. 
In this paper, we will henceforth refer to states of the form Eq. (\ref{eq:pol states}) as {\it fully polarized} states. 

\subsection{Theorems}
We prove two generalizations of Nagaoka's theorem in the SU($n$) Hubbard model. 
The first one is weaker and does not require the connectivity condition. 
However, this version of the theorem does not establish that 
the fully polarized states are the unique ground states. 
The second one is a strict extension of the original theorem in the SU(2) Hubbard model, 
which ensures the uniqueness of the ground states. 

\label{sec:thm}
\begin{thm}
Consider the SU($n$) Hubbard Hamiltonian (\ref{eq:Ham}) with $t_{x,y} \ge 0$, 
$V$ arbitrary, $U = \infty$, and $N_{\rm f}=N_{\rm s}-1$. 
Then among the ground states, there are 
$d_{\rm deg}$ states (see Eq. (\ref{eq:deg})) that are the fully polarized states.  
\label{thm:weak}
\end{thm}
\begin{thm}
Consider the SU($n$) Hubbard Hamiltonian (\ref{eq:Ham}) with $t_{x,y} \ge 0$,
$V$ arbitrary, $U = \infty$, and $N_{\rm f}=N_{\rm s}-1$. 
We further assume that the model satisfies the connectivity condition.  
Then the ground states are fully polarized states and are nondegenerate  
apart from the trivial $d_{\rm deg}$-fold degeneracy due to the SU($n$) symmetry.
\label{thm:strict} 
\end{thm}

\section{Proof}
\label{sec:proof}
In this section we shall prove the theorems.
We first define the basis we work with and clarify the connectivity condition under which Theorem \ref{thm:strict} holds.
Then we prove Theorem \ref{thm:weak} by using the variational principle.
Theorem \ref{thm:strict} is proved as a consequence of the Perron-Frobenius theorem. 

\subsection{Basis states}
In the limit $U\to\infty$, a state with a site occupied by two or more fermions has infinite energy. 
We are interested only in the finite-energy states 
and hence consider the Hilbert space 
\aki{$\mathcal{H}$}
which is spanned by states of the form
\begin{equation}
 \Psi_\Gamma =
  \mathbf{sgn}[x]\left(\prod_{y\in{\La}\bs\{x\}}c_{y,\alpha_y}^\dagger\right)
   \vac,
\label{eq:basis states}
\end{equation}
where $\vac$ is the vacuum (no-particle) state, and  the index  
$\Gamma=(x,\bm{\alpha})$ represents a position $x$ of the hole
(the site without fermion) and a flavor configuration 
$\bm{\alpha}=(\alpha_y)_{y\in\La\bs\{x\}}$ with $\alpha_y=1,2,\dots,n$.
Here, we have assumed that the product is ordered according to an arbitrary
order introduced in $\La$, 
and $\mathbf{sgn}[x]$ takes $-1$ 
if $x$ is an odd-numbered position and $1$ otherwise. 
More explicitly, the states (\ref{eq:basis states}) can be written as
\begin{equation}
\Psi_{\Gamma} = c_{x, \beta} \left(\prod_{y\in{\La}}c_{y, {\tilde \alpha}_y}^\dagger\right)
   \vac,
\label{eq:basis states2}   
\end{equation}
where the flavor configuration ${\tilde {\bm \alpha}}=(\alpha_y)_{y \in \Lambda}$ 
is defined by ${\tilde \alpha}_x = \beta$ for the site $x$ and 
${\tilde \alpha}_y = \alpha_y$ for all $y \in {\La}\bs\{x\}$. 
It is noted that the collection of the states 
\eqref{eq:basis states} as well as \eqref{eq:basis states2} form an orthonormal basis.

Since each number $N_\alpha$ of 
fermions with flavor $\alpha$ is a conserved quantity, 
the Hilbert space $\mathcal{H}$ can be further decomposed into subspaces 
labeled by $(N_\alpha)_{\alpha=1}^n$ with $\sum_{\alpha=1}^n N_\alpha=N_{\rm s}-1$. 
Note that $N_{\rm s}$ denotes the number of sites in $\Lambda$. 
In the following, we denote by $\mathcal{H}[{(N_\alpha)
}]$ the subspace with fixed $(N_\alpha)_{\alpha=1}^n$.

\subsection{Connectivity condition}

Consider the matrix representation of $H$ in the basis defined by Eq. \eqref{eq:basis states}. 
We say that two states labeled by $\Gamma$ and $\Gamma^\prime$  are directly 
connected if $\inprod{\Psi_{\Gamma^\prime}}{H}{\Psi_\Gamma}\ne0$, 
and express this fact by writing $\Gamma\lrarrow\Gamma^\prime$.  
Let $\bm{\alpha}_{y\to x}$
be the flavor configuration on $\Lambda\backslash\{y\}$ obtained from
$\bm{\alpha}$ on $\La\backslash\{x\}$ 
by moving $\alpha_y$ from $y$ to $x$. 
One can easily see that $\Gamma=(x,\bm{\alpha})$ and 
$\Gamma^\prime=(y,\bm{\alpha}^\prime)$ are directly connected if
$t_{x,y}\ne0$ and 
$\bm{\alpha}^\prime=\bm{\alpha}_{y\to x}$, since we have
\begin{equation}
\inprod{\Psi_{\Gamma^\prime}}
{
\left(\sum^n_{\alpha=1} t_{x,y} c^\dagger_{x,\alpha}
 c_{y,\alpha}\right)
}
{
\Psi_{\Gamma}
}=-t_{x,y}.
\label{eq:off diagonal elements}
\end{equation}
Here, we note that the negative sign is attributed 
to the sign factor in the definition of basis states \eqref{eq:basis states}. 
We also note that an off-diagonal matrix element of $H$, i.e.,
\begin{equation} 
\inprod{\Psi_{(y,\bm{\alpha}^\prime)}}{H}{\Psi_{(x,\bm{\alpha})}} = -t_{x,y}~(\le 0),
\end{equation}  
comes solely from the hopping term and 
is non-vanishing only if $(y,\bm{\alpha}^\prime)\lrarrow(x,\bm{\alpha})$.

For two indices $\Gamma$ and $\Gamma^\prime$, 
if there is a sequence of $\bm{\Gamma}=(\Gamma_1,\Gamma_2,\dots,\Gamma_l)$ such that
$\Gamma=\Gamma_1\lrarrow\Gamma_2\lrarrow\cdots
                     \lrarrow\Gamma_l=\Gamma^\prime$,
we say $\Gamma$ and $\Gamma^\prime$ are connected and write 
$\Gamma\leftarrow{\bm{\Gamma}}\rightarrow\Gamma^\prime$.
Then the model is said to satisfy the connectivity condition if all indices corresponding to the basis states 
with common $(N_\alpha)_{\alpha=1}^n$ are connected with each other. 

\textit{Remarks.}
The direct connectivity $(x,\bm{\alpha})\lrarrow(y,\bm{\alpha}_{y\to x})$
implies that the hopping term of the Hamiltonian $H$ transfers the hole
from site $x$ to site $y$ and vice versa.
Similarly, the connectivity $(x,\bm{\alpha})$ and $(y,\bm{\alpha}^\prime)$ implies
that there is a hopping process by which the hole is 
exchanged between $x$ and $y$. 

A simple sufficient condition for the connectivity condition in the SU(2) case was found in Ref.~\cite{Tasaki_1989}. 
In the SU($n$) case, however, it is more difficult to examine the connectivity condition 
for given hopping matrix $t_{x,y}$ and the number of flavors $n$. 
This is because the connectivity condition of our model depends not only on $t_{x,y}$ but also on $n$. 
It may even happen that the model with $n$ less than a certain integer $n_0$ satisfies the
connectivity condition but the one with $n\ge n_0$ does not (e.g. see Ref.~\cite{Miyashita_2009}). 
In Sec. \ref{sec:examples}, we will give several examples where we can verify the connectivity condition explicitly. 

\subsection{Proof of Theorem 1}
Let $\PhiG$ be a ground state of $H$, which is expanded as
$\PhiG=\sum_{\Gamma}\psi_\Gamma\Psi_{\Gamma}$
with certain coefficients $\psi_\Gamma$. 
Let us write $\Psi_{(x,1)}$ for the basis states of
$\mathcal{H}[{(N_\alpha)}]$ with $N_1=N_\mathrm{f}$ 
and $N_{2}=\cdots=N_n=0$ , i.e., 
\begin{equation}
\Psi_{(x,1)} = c_{x,1} \left( \prod_{y \in \Lambda} c^\dagger_{y,1} \right) \Phi_{\rm vac}. 
\end{equation}
Then, consider the trial state 
$\Phi_1=\sum_{x\in\Lambda}\phi_x\Psi_{(x,1)}$
with $\phi_x=(\sum_{\bm{\alpha}}|\psi_{(x,\bm{\alpha})}|^2)^{1/2}$
where the sum is taken over all flavor configurations.

Let us consider the expectation value of $H$ in $\Phi_1$. 
We first find that 
$
\inprod{\Phi_1}{ }{\Phi_1}=\sum_{x}|\phi_x|^2
=\sum_{x}\sum_{\bm{\alpha}}|\psi_{(x,\bm{\alpha})}|^2
=\inprod{\PhiG}{ }{\PhiG}.
$
Next, noting that
$\inprod{\Psi_{(x,1)}}{V(\{n_x \})}{\Psi_{(x,1)}}=
\inprod{\Psi_{(x,\bm{\alpha})}}{V(\{n_x \})}{\Psi_{(x,\bm{\alpha})}}
$
which follows from
\begin{equation}
 n_{x}\Psi_{(y,\bm{\alpha})}=
  \left\{
   \begin{array}{@{\,}ll}
    0                      &\mbox{if $x=y$};\\
    \Psi_{(y,\bm{\alpha})} &\mbox{otherwise},
   \end{array}
	\right.
\end{equation}
for any flavor configurations $\bm{\alpha}$, we get
\begin{eqnarray}
 && \inprod{\Phi_1}{V(\{n_x \})}{\Phi_1}      \nonumber\\
 &=&\sum_{x}|\phi_x|^2\inprod{\Psi_{(x,1)}}{V(\{n_x \})}{\Psi_{(x,1)}}\nonumber\\
 &=&\sum_{x}\sum_{\bm{\alpha}}|\psi_{(x,\bm{\alpha})}|^2
      \inprod{\Psi_{(x,\bm{\alpha})}}{V(\{n_x \})}{\Psi_{(x,\bm{\alpha})}}
                                                                   \nonumber\\
 &=& \inprod{\PhiG}{V(\{n_x \})}{\PhiG}. 
\end{eqnarray}
For the hopping term, we have
\begin{eqnarray}
 && \inprod{\Phi_1}
  {\left(
    \sum_{\alpha=1}^n t_{x,y}c_{x,\alpha}^\dagger c_{y,\alpha}
   \right)}
   {\Phi_1}     
 =-t_{x,y}\phi_{y}\phi_{x} \nonumber\\
 &\le&-t_{x,y}\sum_{\bm{\alpha}}\psi_{(y,\bm{\alpha}_{y\to x})}^\ast \psi_{(x,\bm{\alpha})} \nonumber\\
 &=&
 \lefteqn{ \inprod{\PhiG}
 {\left(
   \sum_{\alpha=1}^n t_{x,y}c_{x,\alpha}^\dagger c_{y,\alpha}
  \right)}
 {\PhiG}}.     
\end{eqnarray}
Here, we have used the Schwarz inequality
\begin{eqnarray}
\lefteqn{
  \sum_{\bm{\alpha}}\psi_{(y,\bm{\alpha}_{y\to x})}^\ast
                    \psi_{(x,\bm{\alpha})}
  }\nonumber\\
&\le&
\left(\sum_{\bm{\alpha}}|\psi_{(y,\bm{\alpha}_{y\to x})}|^2 \right)^\frac{1}{2}
\left(\sum_{\bm{\alpha}}|\psi_{(x,\bm{\alpha})}|^2
\right)^\frac{1}{2}
\nonumber\\
&=&\phi_y\phi_x.
\end{eqnarray}
As a consequence of the relations obtained above,
we find 
\begin{equation}
\frac{\inprod{\Phi_1}{H}{\Phi_1}}{\inprod{\Phi_1}{ }{\Phi_1}}
\le
\frac{\inprod{\PhiG}{H}{\PhiG}}{\inprod{\PhiG}{ }{\PhiG}},
\end{equation}
which shows that $\Phi_1$ is also a ground state. 
Then the theorem follows by taking $\Phi_1$ as one of the ground states and using 
the global SU($n$) symmetry discussed in Sec. \ref{sec: full pol}. 

\begin{figure*}[tb]
\begin{center}
 \includegraphics[width=.7\textwidth]{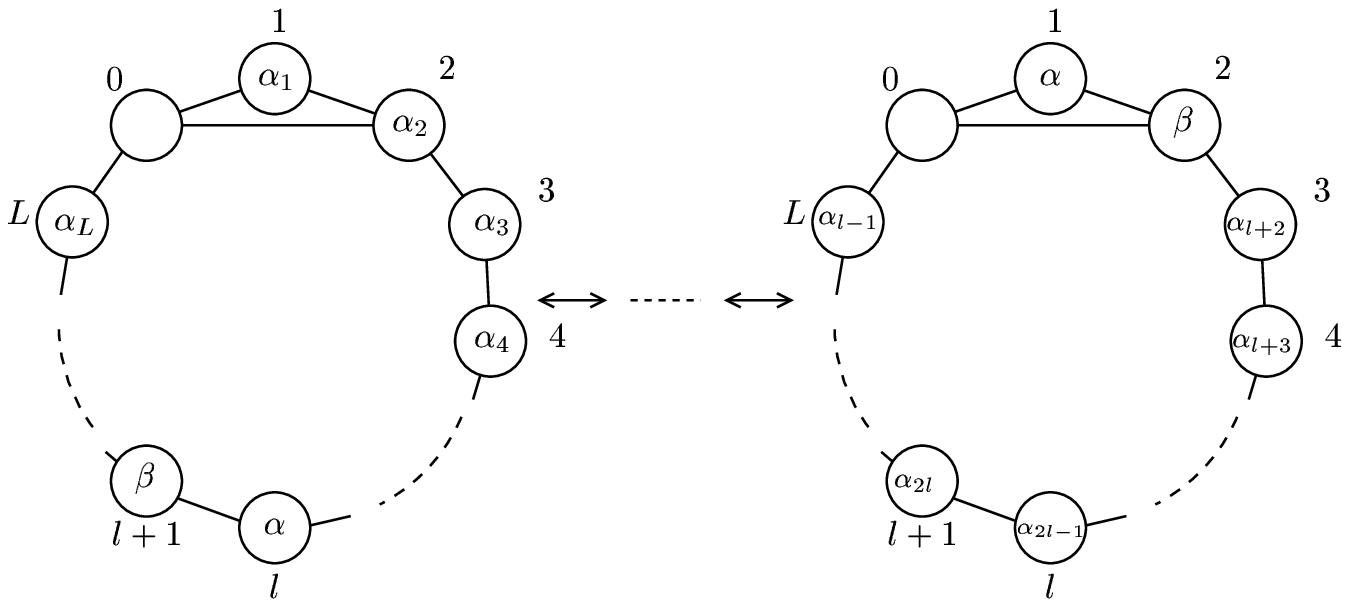}
\end{center}
\caption{By letting the hole move around the chain in the clockwise direction
$l-1$ times, one can bring the fermions from sites $l$ and $l+1$ to sites $1$ and $2$, respectively. 
}
\label{fig:chain}
\end{figure*}
\begin{figure}[htb]
\begin{center}
 \includegraphics[width=.9\columnwidth]{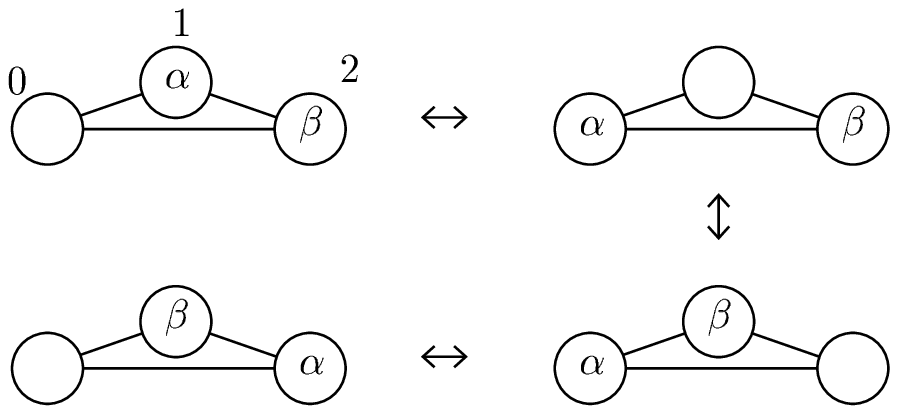}
\end{center}
\caption{The flavors at sites $1$ and $2$ are exchanged when the hole hops around the loop 
$\{0,1,2\}$ in the clockwise direction once. 
}
\label{fig:triangle}
\end{figure}

\subsection{Proof of Theorem 2}
Let us fix $(N_\alpha)_{\alpha=1}^n$ and consider the matrix
representation of $H$ for the basis states $\Psi_\Gamma$ in $\mathcal{H}[(N_\alpha)]$.
Then, as we noted below \eqref{eq:off diagonal elements}, 
all the off-diagonal matrix elements are non-positive.
Furthermore, the connectivity condition ensures that the matrix is indecomposable.
Therefore, the Perron-Frobenius theorem is applicable to the present
matrix, implying that the lowest-energy state in the subspace 
$\mathcal{H}[(N_\alpha)]$ is unique.

This fact together with the statement of Theorem 1 will complete the proof of Theorem 2. 
However, we give here a simple proof which relies only on the Perron-Frobenius theorem. 
We use another consequence of the Perron-Frobenius theorem: the lowest-energy state in 
each subspace is given by a certain linear combination of all $\Psi_\Gamma$ 
with positive coefficients. 

Let $\Phi_{1,\mathrm{G}}$ be the lowest-energy state
with energy $E$ in the subspace where all fermions have flavor~1. 
Using the operators $F^{\alpha,\beta}$ in Eq. (\ref{eq:gene}), 
one can construct the state with the same energy $E$ in the subspace $\mathcal{H}[(N_\alpha)]$ as
\begin{equation}
(F^{n,1})^{N_n}\cdots(F^{3,1})^{N_3}(F^{2,1})^{N_2}\Phi_{1,\mathrm{G}}.
\label{eq:ground states}
\end{equation}
Here note that the state \eqref{eq:ground states} 
can be expanded in terms of $\Psi_\Gamma$ with positive coefficients. 
Since the lowest-energy state in $\mathcal{H}[(N_\alpha)]$ also has positive coefficients,
it is not orthogonal to the state \eqref{eq:ground states}.
Therefore, the state \eqref{eq:ground states} is exactly 
the lowest-energy state in the subspace $\mathcal{H}[(N_\alpha)]$.  
This implies that the ground state is unique apart from
the degeneracy due to the SU($n$) symmetry. 

\section{Examples}
\label{sec:examples}
In this section we give three examples where the connectivity condition 
can be verified explicitly. 
Before proceeding, let us introduce some terminology. 
A pair $\{x,y\}$ of sites in $\Lambda$ is called a {\it bond} if $t_{x,y}\ne0$.
By a {\it path} from $x$ to $y$, we mean an ordered set $\{x_1,x_2,\dots,x_l\}$ of $l$~sites
such that $x_1=x,~x_l=y$ and $\{x_{m},x_{m+1}\}$ is a bond for all $m=1,\dots,l-1$.   
A path $\{x_1,x_2,\dots,x_l\}$ is called a {\it loop} if $\{x_1,x_l\}$ is a bond.
A lattice $\Lambda$ is said to be connected
if one can find a path for any two sites $x$ and $y$ in $\La$.
A connected lattice $\Lambda$ is said to be two-fold connected,
if one cannot make it disconnected by removing a single site. 

The connectivity of $\La$ implies that one can bring the hole to 
any site in $\La$ by successive hops. It should be noted, however, that 
the connectivity of $\La$ itself does not necessarily mean the connectivity 
of the model. This is because the motion of the hole around the lattice 
may not generate all the configurations $\Gamma=(x,{\bm \alpha})$ with common $(N_\alpha)^n_{\alpha=1}$. 

\subsection{Closed chain with one next nearest neighbor bond}
The first example is the SU($n$) Hubbard Hamiltonian on a closed chain, i.e., 
the one-dimensional lattice $\La=\{0,1,\dots,L\}$ with bonds $\{l,l+1\}~(l=0,1,\dots,L-1)$ and $\{L,0\}$.
It is, furthermore, assumed that there is one next nearest neighbor bond, say $\{0,2\}$. 
Let us take a fixed $(N_\alpha)_{\alpha=1}^n$ and examine the connectivity of 
indices $\Gamma=(x,\bm{\alpha})$ in this subspace. 
Since the lattice $\La$ is obviously two-fold connected, one can bring the hole to any site. 
Thus, one can fix the location of the hole, say $0$, and see whether the motion of the hole 
generate all the flavor configurations $(0,\bm{\alpha})$ in the subspace. 

For this purpose, we show below that, for any bond $\{x,y\}$,
$(0,\bm{\alpha})$ and $(0,\bm{\alpha}_{x\lrarrow y})$ are connected,
where $\bm{\alpha}_{x\lrarrow y}$ is the flavor configuration obtained from $\bm{\alpha}$
by switching $\alpha_x$ and $\alpha_y$.
This property implies that one can generate any flavor configuration $\bm{\alpha}^\prime$ 
from $\bm{\alpha}$ by successively switching the flavors on a pair of neighboring sites in an appropriate way. 
Let us assume that $\alpha_{l}=\alpha$ and $\alpha_{l+1}=\beta$.
By letting the hole move around the closed chain in the clockwise direction $l-1$
times, we first obtain the flavor configuration $\bm{\alpha}^\prime$ in which 
$\alpha_1^\prime=\alpha$ and $\alpha_2^\prime=\beta$ (see Fig.~\ref{fig:chain}).
Next we let the hole move around the triangle loop $\{0,1,2\}$ in the clockwise
direction once, by which the flavor $\alpha_1^\prime$ is exchanged with
$\alpha_2^\prime$ (see Fig.~\ref{fig:triangle}).
Finally, 
by letting the hole move along the chain in the opposite(counter-clockwise) direction
$l-1$ times,
we get the flavor configuration $\bm{\alpha}^{\prime\prime}$ with 
$\alpha_{l}^{\prime\prime}=\beta,~\alpha_{l+1}^{\prime\prime}=\alpha$ 
without changing the flavor configuration outside $\{ l, l+1 \}$. 
This proves that the model satisfies the connectivity condition. 
 
A few comments are in order: the presence of the single next-nearest-neighbor bond 
changes drastically the nature of ground states. In fact, in the purely one-dimensional chain 
in which the connectivity condition is not satisfied, all the flavor configurations become 
degenerate at $U = \infty$ in the same way as in the standard SU(2) case~\cite{Ogata_Shiba_1990}. 
This degeneracy is removed if $U < \infty$ and the ground states form a unique 
antisymmetric multiplet, which can be proved using the Perron-Frobenius argument~\cite{Hakobyan_2010}. 
In particular, the ground state is a unique SU($n$) singlet if the number of fermions is a multiple of $n$. 

\subsection{Two-fold connected lattice with triangle loops}
The second example is a class of models defined on two-fold connected lattices 
containing at least one triangle loop. Typical examples are the triangular and 
kagome lattices. 
Let us show that the model of this class satisfies the connectivity condition.
Let $\{u,v,w\}$ be a triangle loop, and let $\{x,y\}$ be an arbitrary bond in the lattice $\Lambda$.
Since $\Lambda$ is connected, we can find a path from $x$ to $u$
and another path from $y$ to $v$.
Suppose that these two paths share the same site $z$ 
and one cannot find other paths that do not intersect. 
Then, the removal of the site $z$ makes $\Lambda$ disconnected.  
This contradicts the fact that $\La$ is two-fold connected. 
Thus one can always find two non-intersecting paths; 
one from $x$ to $u$ and the other from $y$ to $v$. 
It follows from this that there is a loop containing
two bonds $\{u,v\}$ and $\{x,y\}$ in $\Lambda$. 
Noting that $\{u,v\}$ belongs to the triangle loop,
we can repeat the same argument as in the previous example,
which concludes that the connectivity condition is satisfied in the model. 

\subsection{Lattice consisting of loops of 4 sites}
\label{sec:loop4}
In the previous two examples, the existence of a loop of three sites plays an important
role in establishing the connectivity condition. 
It is then natural to ask whether the connectivity condition is satisfied or not 
in a model on a lattice without any triangle. 
However, it seems difficult to answer this question in a general setting. 
Here, we instead give a concrete example: the model on a lattice 
consisting only of loops of length four. 
Our lattice $\La$ can be constructed as follows. 
Let $C_1$ be a loop of four sites having four bonds.
Let $C_l$ with $l=2,\dots,L$ be identical copies of $C_1$. 
We construct $\La_2$ by adding $C_2$ to $\La_1=C_1$ in such a way that
$\La_1$ and $C_2$ share one bond. 
Similarly, we construct $\La_l$ by adding $C_l$ to $\La_{l-1}$ in such a way
that  $C_l$ shares at least one bond with $\La_{l-1}$ 
and shares at most one bond with each $C_{l^\prime}~(l^\prime=1,\dots,l-1)$ 
that forms $\La_{l-1}$. 
Our lattice is then obtained as $\La = \La_L$. 
Typical examples are square and cubic lattices. 

In the following, we shall prove that the model on $\La$ with $N_{\rm s}\ge n+2$ satisfies 
the connectivity condition, i.e., any pair of indices $\Gamma=(z,\bm{\alpha})$ and $\Gamma^\prime=(z^\prime,\bm{\alpha}^\prime)$ with common $(N_\alpha)_{\alpha=1}^n$ are connected. 
To see this, let us consider the subspace with fixed $(N_\alpha)_{\alpha=1}^n$.
Since we have assumed that $N_{\rm s}\ge n+2$, the fermion number
$N_\mathrm{f}$ satisfies $N_{\mathrm{f}}=N_{\rm s}-1\ge n+1$.
It is thus ensured that there is a flavor $\sigma$ such that $N_\sigma\ge2$. 
The presence of two fermions with flavor $\sigma$ plays an important role in the proof below. 
As in the previous examples, it suffices to show that $(z,\bm{\alpha})$ and $(z,\bm{\alpha}_{x\lrarrow y})$ are connected for any bond $\{x,y\}$ since the lattice $\La$ is obviously two-fold connected. 

\begin{figure}[tbh]
\begin{center}
\includegraphics[width=.80\columnwidth]{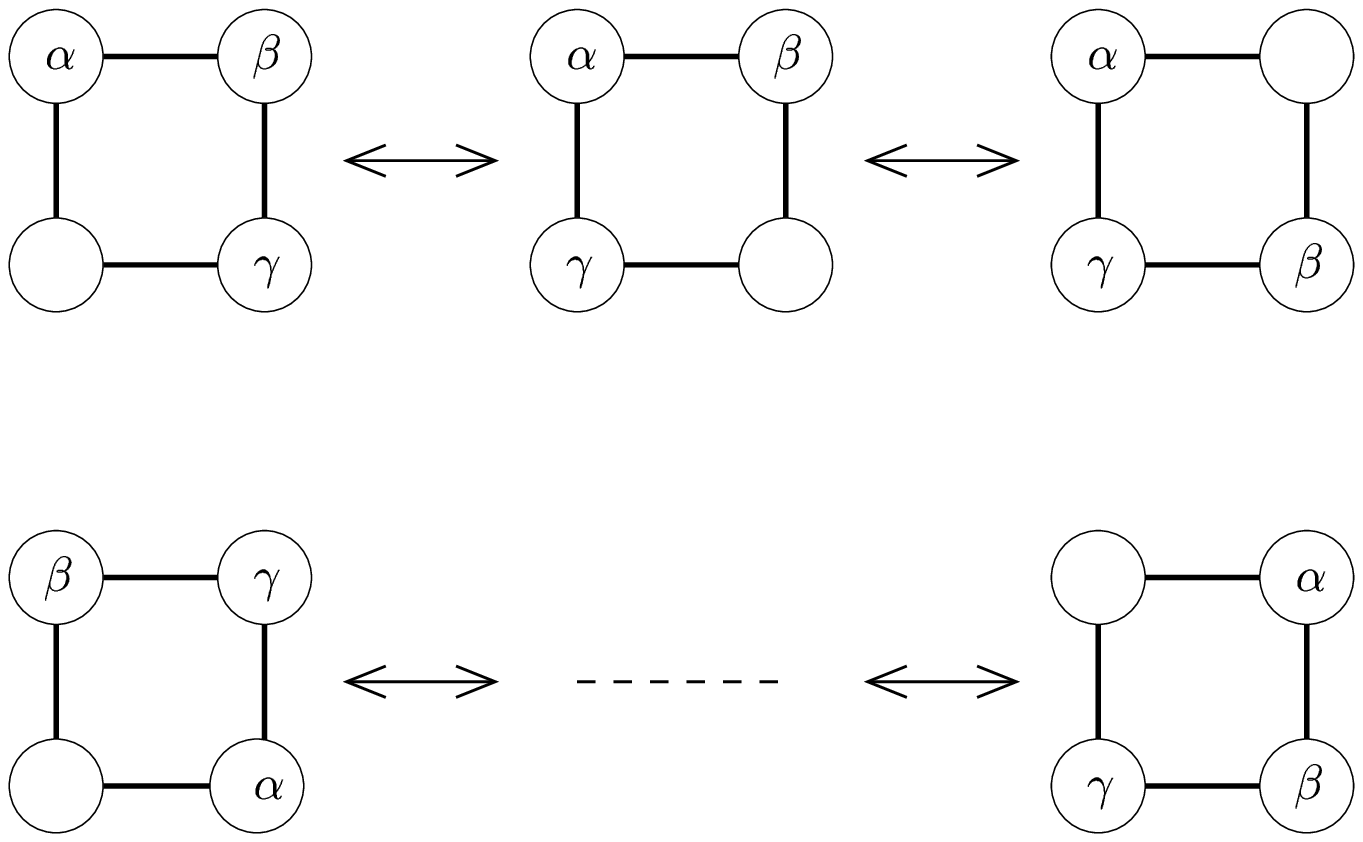}
\end{center}
\caption{Three fermions and the hole in a loop $C_l$. 
The motion of the hole does not change the order of flavors in the clockwise (counter-clockwise) direction. 
}
\label{fig:motion of the hole}
\end{figure}
\begin{figure}[tbh]
\begin{center}
\includegraphics[width=\columnwidth]{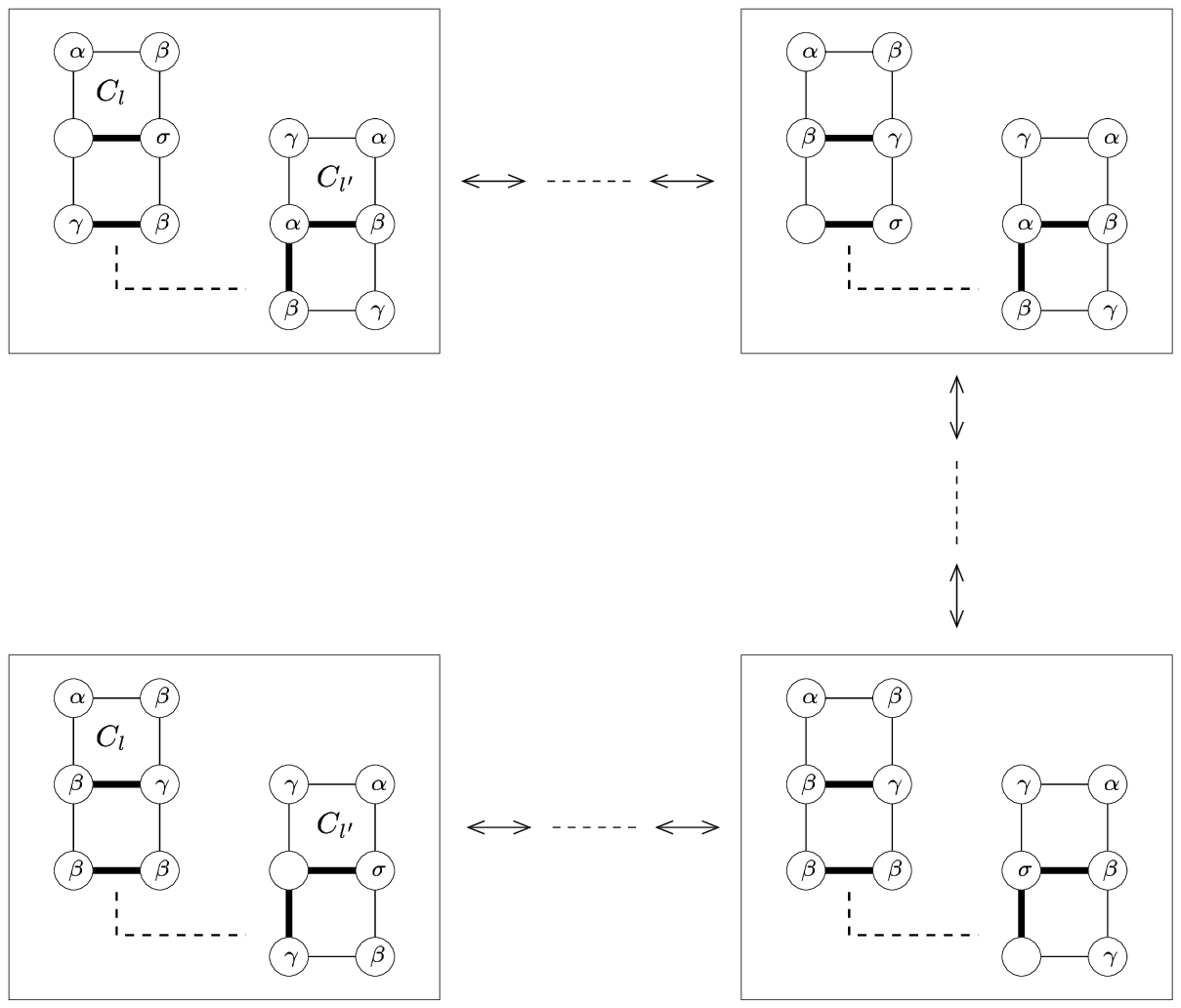}
\end{center}
\caption{
A pair of the hole and a fermion with flavor $\sigma$ in loop $C_l$ can
be moved to any other loop $C_{l^\prime}$ through shared bonds 
emphasized by thick lines.}
\label{fig:motion of the pair}
\end{figure}
Before proceeding, we give two remarks on the connectivity and the
motion of the hole: 
(i) in a loop $C_l$ containing three fermions and the hole, the motion of the hole 
does not change the order of flavors in the clockwise (counter-clockwise) direction 
(see Fig.~\ref{fig:motion of the hole}).
(ii) One can bring a pair of the hole and a fermion with a fixed flavor 
from a loop $C_l$ to any other loop $C_{l^\prime}$. 
This is confirmed by the first property (i) and the construction of $\La$. 
In fact, there is a sequence of $C_l=C_{l_1}, C_{l_2}, \dots,
C_{l_k}=C_{l^\prime}$ such that $C_{l_{i}}$ and $C_{l_{i+1}}$ share a bond,
and, by occupying these shared bonds, a pair of the hole and a fermion can move from $C_l$ to
$C_{l^\prime}$ (see Fig.~\ref{fig:motion of the pair}). 

Let us prove the connectivity of $(z,\bm{\alpha})$ and
$(z,\bm{\alpha}_{x\lrarrow y})$ for an arbitrary bond $\{x,y\}$
by explicitly constructing a sequence $\bm{\Gamma}$ such that
$(z,\bm{\alpha})\leftarrow \bm{\Gamma} \rightarrow (z,\bm{\alpha}_{x\lrarrow y})$.
Here we assume that $\alpha_x=\alpha$ and $\alpha_y=\beta$.
From the construction of $\La$, it is possible to find a loop $C_l=\{x,y,u,v\}$ and a path 
from $z$ to $v$ that contains neither $x$ nor $y$. 
This means that we have $\bm{\Gamma}^{(1)}$ such that
$(z,\bm{\alpha})\leftarrow \bm{\Gamma}^{(1)} \rightarrow
(z^{(1)},\bm{\alpha}^{(1)})$ where $z^{(1)}(\ne x,y)$ is a site in
$C_l$ and $\bm{\alpha}^{(1)}$ is a certain flavor configuration.
We recall that there is a flavor $\sigma$ with $N_{\sigma}\ge 2$.
By using two fermions with flavor $\sigma$, we will construct a process
in which $\alpha_x$ and $\alpha_y$ are switched.

\begin{figure*}[htb]
\begin{center}
\includegraphics[width=.7\textwidth]{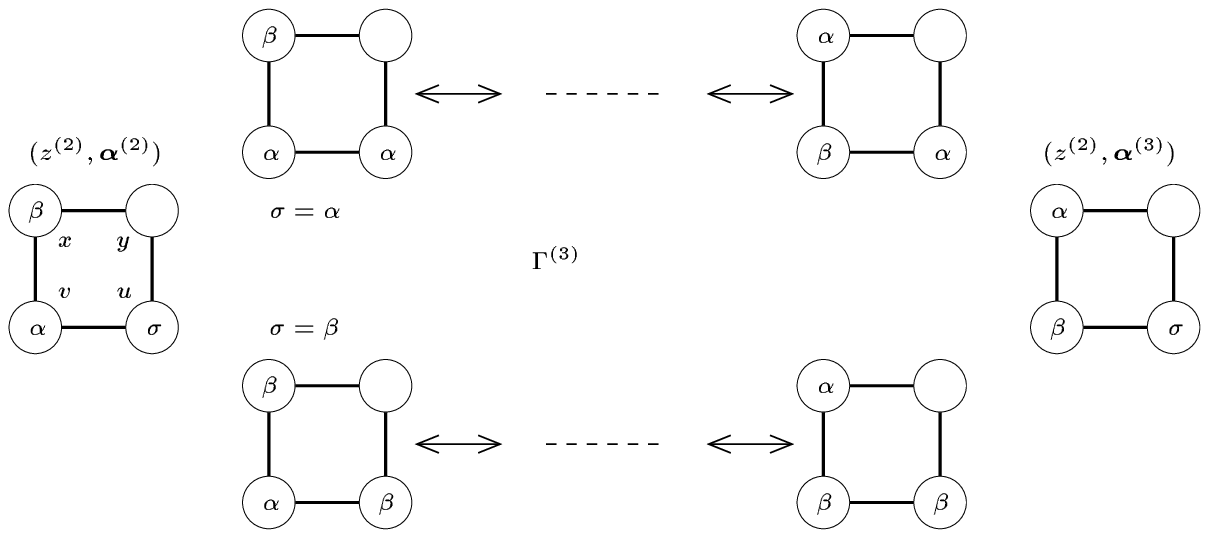}
\end{center}
\caption{
$\bm{\Gamma}^{(3)}$ connecting $(z^{(2)},\bm{\alpha}^{(2)})$ to
$(z^{(2)},\bm{\alpha}^{(3)})$. It is assumed that $\{y,u\}$ is a shared
 bond, so that fermions with flavor $\alpha$ and $\beta$ were transferred
 in the process $\bm{\Gamma}^{(2)}$. In $(z^{(2)},\bm{\alpha}^{(2)})$,
 the position of the hole is $z^{(2)}=y$, and the local flavor
 configuration is $\alpha_x^{(2)}=\beta$, $\alpha_v^{(2)}=\alpha$,
 and $\alpha_u^{(2)}=\sigma$,
while, in $(z^{(2)},\bm{\alpha}^{(3)})$,
 the local flavor configuration is changed to 
  $\alpha_x^{(3)}=\alpha$, $\alpha_v^{(3)}=\beta$, and
 $\alpha_u^{(3)}=\sigma$.
 $\bm{\Gamma}^{(3)}$ exchanges the flavors on the bond $\{ v,x \}$ 
 without changing the configuration outside $\{ v,x \}$
}
\label{fig:Gamma three}
\end{figure*}

Now suppose that one of the fermions with flavor $\sigma$ is in
$C_{l^\prime}$.
We can find a sequence of shared bonds from $C_l$ to
$C_{l^\prime}$, as mentioned above.
If $x$ or $y$ $\in C_l$ (or both of them) touches the shared bond we transfer 
the two fermions on $\{ x,y \}$ to the sites on $\{ u,v \}$. 
Then we can bring the hole in $C_l$ to $C_{l^\prime}$ through
sites touching the shared bonds,
and from (ii) we can bring back the pair of the hole and the fermion with flavor
$\sigma$
from $C_{l^\prime}$ to $C_{l}$.
Thus we can find $\bm{\Gamma}^{(2)}$ which connects
$(z^{(1)},\bm{\alpha}^{(1)})$
to $(z^{(2)},\bm{\alpha}^{(2)})$ where $z^{(2)}$ is a site in
$C_l=\{x,y,u,v\}$, and $\bm{\alpha}^{(2)}$ is a certain flavor configuration.
We note that now there are three fermions with flavor $\alpha$, $\beta$ and $\sigma$
in $C_l$.
(See Fig.~\ref{fig:Gamma three} for an example of a local flavor configuration 
$(\alpha_x^{(2)},\alpha_y^{(2)},\alpha_u^{(2)},\alpha_v^{(2)})$.)

\begin{figure*}[htb]
\begin{center}
 \includegraphics[width=.9\textwidth]{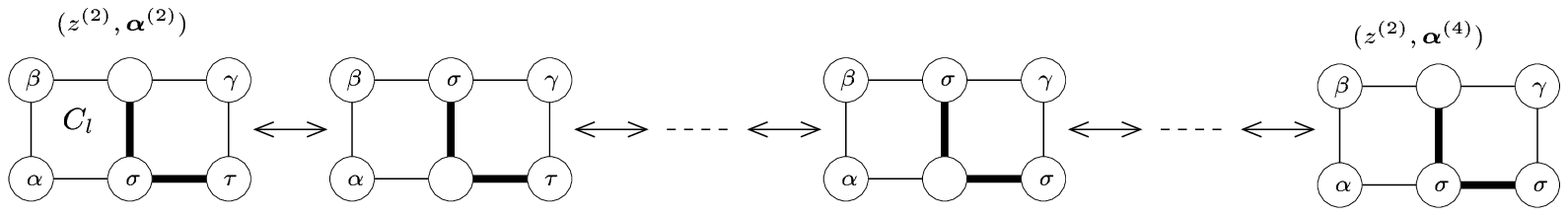}
\end{center}
\caption{$\bm{\Gamma}^{(4)}$ connecting $(z^{(2)},\bm{\alpha}^{(2)})$ to
$(z^{(2)},\bm{\alpha}^{(4)})$. The thick lines represent shared bonds.}
\label{fig:Gamma four}
\end{figure*}

If $\alpha$ or $\beta$ equals $\sigma$, 
as depicted in Fig.~\ref{fig:Gamma three},
we can find the sequence
$\bm{\Gamma}^{(3)}$
which connects $(z^{(2)},\bm{\alpha}^{(2)})$ to
$(z^{(2)},\bm{\alpha}^{(3)})$, 
where $\bm{\alpha}^{(3)}$ 
is obtained from $\bm{\alpha}^{(2)}$ 
by switching two flavors $\alpha$ and $\beta$ on a bond in $C_l$.
Then, by tracing the motion of the pair of the hole and 
the fermion with flavor $\sigma$ backwards, we find the sequence $\overline{\bm{\Gamma}^{(2)}}$ 
such that $(z^{(2)},\bm{\alpha}^{(3)})\leftarrow \overline{\bm{\Gamma}^{(2)}} 
 \rightarrow (z^{(1)},\bm{\alpha}_{x\lrarrow y}^{(1)})$. 
Similarly, we can find the sequence 
$(z^{(1)},\bm{\alpha}_{x\lrarrow y}^{(1)}) \leftarrow \overline{\bm{\Gamma}^{(1)}} \rightarrow (z,\bm{\alpha}_{x\lrarrow y})$ 
by tracing the hole motion backwards. 
A desired sequence $\bm{\Gamma}$ is then obtained by setting 
$\bm{\Gamma}=(\bm{\Gamma}^{(1)},\bm{\Gamma}^{(2)},\bm{\Gamma}^{(3)},
                   \overline{\bm{\Gamma}^{(2)}},\overline{\bm{\Gamma}^{(1)}})$.
\begin{figure*}[htb]
\begin{center}
 \includegraphics[width=.8\textwidth]{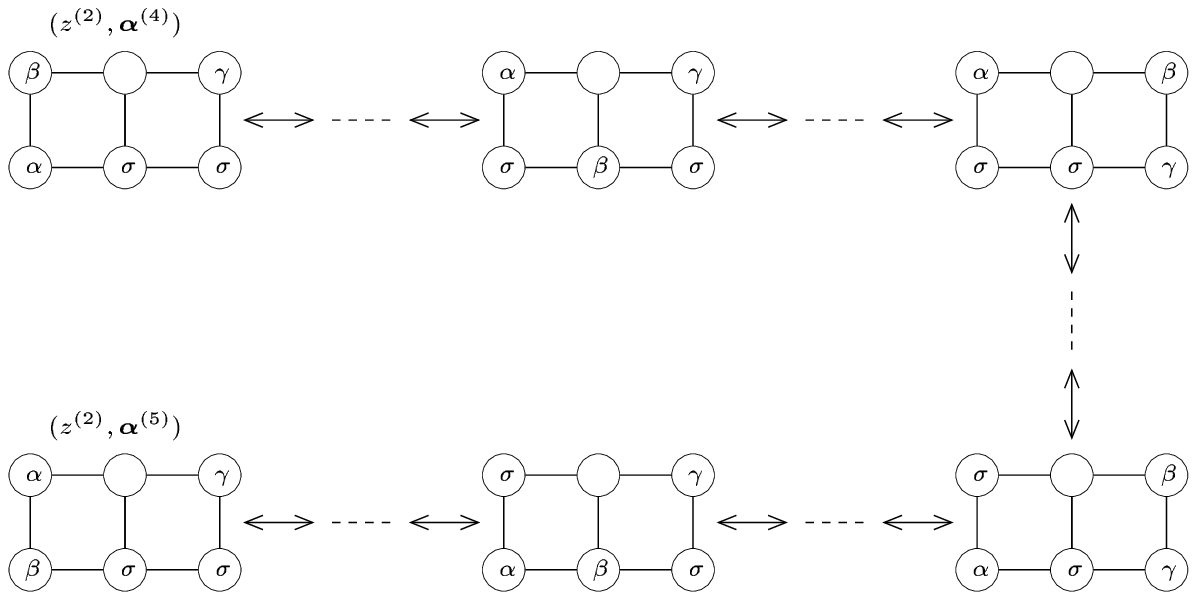}
\end{center}
\caption{$\bm{\Gamma}^{(5)}$ connecting $(z^{(2)},\bm{\alpha}^{(4)})$ to
$(z^{(2)},\bm{\alpha}^{(5)})$.}
\label{fig:Gamma five}
\end{figure*}

If neither $\alpha$ nor $\beta$ equals $\sigma$, we have to find
another fermion with flavor $\sigma$ in a certain loop. 
Repeating the above argument, 
we can find $\bm{\Gamma}^{(4)}$ which connects
$(z^{(2)},\bm{\alpha}^{(2)})$
to $(z^{(2)},\bm{\alpha}^{(4)})$ where the local flavor configuration
of $\bm{\alpha}^{(4)}$ is indicated in Fig.~\ref{fig:Gamma four}.
Then, as shown in Fig.~\ref{fig:Gamma five}, we have 
$(z^{(2)},\bm{\alpha}^{(4)})\leftarrow
\bm{\Gamma}^{(5)}\rightarrow (z^{(2)},\bm{\alpha}^{(5)})$
where $\bm{\alpha}^{(5)}$ 
is obtained from $\bm{\alpha}^{(4)}$ 
by switching two flavors $\alpha$ and $\beta$ on a bond in $C_l$. 
By introducing $\overline{\bm{\Gamma}^{(4)}}$ in the same manner as above, 
we find  
\begin{eqnarray*}
&(z^{(2)},\bm{\alpha}^{(5)})\leftarrow
(\overline{\bm{\Gamma}^{(4)}},\overline{\bm{\Gamma}^{(2)}})
\rightarrow& 
(z^{(1)},\bm{\alpha}_{x\lrarrow y}^{(1)})\\
&&\leftarrow
\overline{\bm{\Gamma}^{(1)}}
\rightarrow
(z,\bm{\alpha}_{x\lrarrow y}),
\end{eqnarray*}
which gives a desired sequence 
\[
\bm{\Gamma}=(\bm{\Gamma}^{(1)},\bm{\Gamma}^{(2)},\bm{\Gamma}^{(4)},\bm{\Gamma}^{(5)},
                \overline{\bm{\Gamma}^{(4)}},\overline{\bm{\Gamma}^{(2)}},\overline{\bm{\Gamma}^{(1)}}).
\]

\section{Conclusion and outlook}
\label{sec:summary}
We have presented an extension of Nagaoka's theorem to the infinite-$U$ 
Hubbard model with SU($n$) symmetry. 
Similar to the SU(2) case, for the model with one hole, we found that 
(i) the fully polarized (Nagaoka) states analogous to the ferromagnetic states are ground states; 
(ii) these Nagaoka states are the only possible ground states if the connectivity condition is satisfied. 
However, unlike the SU(2) case, here it is not easy to verify the connectivity condition for given $n$ and lattice structure.  
A simple sufficient condition for the connectivity condition is that the lattice 
contains at least one triangle loop. We also found a class of lattices consisting 
of loops of length four that satisfy the connectivity condition when the total number of sites $N_{\rm s}$ is large enough ($N_{\rm s} \ge n+2$). 
Examples include hypercubic lattices in $d \ge 2$ dimensions. 

An interesting question is whether the Nagaoka states discussed in the present paper can be detected experimentally. 
For the SU(2) case, a controllable scheme to detect Nagaoka ferromagnetism 
in optical superlattices has been proposed~\cite{Stecher_Demler_2010}. 
In this scheme, we need to prepare an array of isolated plaquettes each of 
which consists of four lattice sites with three fermions. 
The Nagaoka transition and the variation of the total spin in the ground state 
can then be probed using a band mapping analysis~\cite{Greiner_2001} after switching off the superlattice potential. 

We now extend the above scheme to the SU($n$) case with $n > 2$. 
Since positive hopping amplitudes are more difficult to achieve experimentally, 
we focus on the case where the lattice is bipartite and is formed by loops of length four (see Sec. \ref{sec:loop4}). 
In this case, one needs to prepare an array of clusters each of which 
consists of more than $n+2$ sites. 
A simple choice is to take each cluster as a cube with eight lattice sites, 
which implies that $n \le 6$. 
Fortunately, preparation of the SU(6) Hubbard system is feasible in current experiments 
with ${}^{173}$Yb atoms~\cite{Takahashi_group_2012}. 
Furthermore, an array of cubes can be easily created by superimposing 
optical lattices with different periodicities. 
We thus expect that the onset of the Nagaoka states in the SU($n$) Hubbard 
model can be studied experimentally with currently-available techniques.

Finally, we remark on the instability of the Nagaoka states. 
In the standard SU($2$) case, one can prove that the Nagaoka states are 
{\it not} the ground states for large enough densities of holes by constructing 
a variational state with one overturned spin which has a lower energy than 
the Nagaoka states~\cite{Shastry_1990, Basile_1990, Suto_1991}. 
For the model on a square lattice, for example, an extension of the earlier work 
yields the best estimate of the critical hole density $\delta_{\rm cr}=0.251$,  
above which the Nagaoka state is unstable~\cite{Muller_1996}. 
The same variational argument applies to the SU($n$) case, because the model 
in the subspace with $N_1+N_2=N_{\rm f}$ and $N_3=\cdots = N_{n}=0$ is described 
by the SU(2) Hubbard Hamiltonian. Thus we see that the SU($n$) Nagaoka 
states are also unstable with respect to the finite concentration of holes, 
implying the importance of the precise control of hole densities 
in a possible realization of the fully polarized states.  

\acknowledgments

We thank Leon Balents and Alexey Gorshkov for stimulating discussions. 
H.K. was supported in part by Grant-in-Aid for Young Scientists (B) (Grant No. 23740298).

\end{document}